\documentclass[11pt,a4paper,fleqn]{article}

\usepackage[longnamesfirst]{natbib}
\usepackage{graphicx}        
\usepackage[bottom]{footmisc}

\usepackage{amsbsy}
\usepackage{amsfonts}
\usepackage{amsmath}
\usepackage{enumerate}
\usepackage{mathrsfs} 
\usepackage{hyperref}

\newcommand{\E}{\mathrm{\mathbf E}}
\newcommand{\bOne}{\boldsymbol{1}}
\newcommand{\bZero}{\boldsymbol{0}}
\newcommand{\ba}{\boldsymbol{a}}
\newcommand{\bb}{\boldsymbol{b}}
\newcommand{\be}{\boldsymbol{e}}
\newcommand{\bx}{\boldsymbol{x}}
\newcommand{\bmu}{\boldsymbol{\mu}}
\newcommand{\hS}{\hat{S}}
\newcommand{\hV}{\hat{V}}
\newcommand{\bX}{\boldsymbol{X}}
\newcommand{\hSigma}{\hat\Sigma}
\newcommand{\dsX}{\mathbb{X}}
\newcommand{\Ss}{\mathscr{S}} 
\def\Sp{\Ss^+_p}
\def\R{\mathbb{R}}
\DeclareMathOperator*{\argmin}{arg\,min}
\DeclareMathOperator*{\trace}{trace}
\newcommand{\mI}{\mathcal{I}}
\newcommand{\mK}{\mathcal{K}}
\newcommand{\mM}{\mathcal{M}}
\newcommand{\vv}{\rm vec}

\begin{document}

\title{Robustly fitting Gaussian graphical models: \\
       the {R}-package robFitConGraph}
\author{Daniel Vogel\thanks{Corresponding author: daniel.vogel@tu-dortmund.de}, \ Stuart J.~Watt, and Anna Wiedemann}
\date{December 1, 2022}

\maketitle

\begin{abstract}
\small
\noindent
This paper gives a tutorial-style introduction to the R-package robFitConGraph, which provides a robust goodness-of-fit test for Gaussian graphical models. Its use is demonstrated at a data example on music performance anxiety, which also illustrates \emph{why} one would want to fit a Gaussian graphical model -- and why one should do so robustly. The underlying statistical theory is briefly explained.

The paper describes package version 0.4.1, available on CRAN from December 1, 2022. See \url{https://CRAN.R-project.org/package=robFitConGraph}.
\end{abstract}


\section{Introduction}

The first two Sections 1 \emph{Introduction} and 2 \emph{A Case Study} are intended for a general audience, assuming neither deeper familiarity with graphical models nor robustness. Section 3 \emph{Background and Theory} discusses some aspects in detail.

\subsection{Gaussian graphical modeling}
\label{sec:ggm}

Graphical models are an important tool for analyzing the dependence structure of several variables. Gaussian graphical models are employed for continuously distributed data, where a multivariate normal, or Gaussian, distribution is adequate. 

A graph encodes the dependence structure of a random vector $\bX = (X_1,\ldots, X_p)$ as follows: The nodes represent the individual variables, and an edge between two nodes represents partial correlation, or the absence of an edge zero partial correlation. Partial correlation is a measure for conditional dependence between two variables, conditional on all remaining variables, which, in a very loose sense, can be understood as a measure for dependence \emph{not} explained by joint dependence on other variables. 
As soon as more than two variables are considered, conditional dependence is arguably more important than marginal dependence: It suggests and helps to verify causal relationships. While ice cream sales and criminal assault rates per day -- at any given place in the temperate climate zone -- are certainly positively correlated, their conditional independence given a suitable mediator variable, such as outside mean temperature, provides evidence for why this may be so. 

Graphical modeling refers to the statistical task of finding an appropriate graph that describes the dependence structure of a given data set, i.~e., identifying all zero partial correlations. The aim is to find a parsimonious graph, i.~e., one with few edges, which does not contradict the data. A full graph, with all edges present, means no restriction and contains no structural information. A completely empty graph, with no edges at all, means all variables are independent. 

Let $\Sigma$ denote the covariance matrix of $\bX$. Throughout, we assume that all second moments of $\bX$ are finite, and that furthermore $\Sigma$ is positive definite and thus may be inverted to yield the \emph{concentration matrix} or \emph{precision matrix} $K = \Sigma^{-1}$. The assumption of strict positive definiteness is a mild one and is equivalent to the probability mass not being concentrated on a lower-dimensional affine linear subspace of $\mathbb{R}^p$. We make the same assumption on the data
points $\dsX_n = (\bx_1, \ldots \bx_n)^\top$, and hence the sample covariance matrix $\hat\Sigma_n$ is positive definite and the sample concentration matrix 
$\hat{K}_n = \hat\Sigma_n^{-1}$ exists.

The basis for Gaussian graphical modeling is the following characterization: The variables $X_i$ and $X_j$ are partially uncorrelated given all remaining $p-2$ variables if and only if $K_{i,j} = 0$. The \emph{partial correlation} $p_{i,j}$ of $X_i$ and $X_j$ given the remaining components of $\bX$ is defined as the correlation between the residuals of $X_i$ and $X_j$ when regressing both on the remaining components. Some matrix calculus yields that
\begin{equation} \label{eq:partial}
	p_{i,j} = -\frac{K_{i,j}}{\sqrt{K_{i,i} K_{j,j}}}, 
\end{equation}
see, e.~g., \citet[][Chapter 5]{Whittaker1990}. Thus the pairwise partial correlations are obtained from the inverse covariance matrix $K$ in a very similar fashion as the pairwise correlations are obtained from the covariance matrix $\Sigma$ itself. The only difference is the minus sign. Consequently, an absent edge in the graph means a zero entry in $K$, and finding the graph for given data comes down to finding the zero-pattern in the inverse of the true covariance matrix $\Sigma$. 
%
%
Three basic sub-tasks of graphical modeling can be identified:
\begin{enumerate}[({T}1)]
\item Finding an appropriate graph.
\item Determining if a given graph fits the data. 
\item Estimating the (remaining) partial correlations under a given graph structure. 
\end{enumerate}

Whether (T1) or (T2) is considered more important may be debatable and mainly depends on whether one pursues an explorative or an inferential analysis. Task (T3) may appear of lesser importance, but it is intrinsically linked to (T2).
We briefly outline how these tasks are approached, in reverse order, starting in with (T3).

\paragraph{Task (T3)}
We require some mathematical notation. 
Define a graph $G = (V,E)$ as a set of vertices $V = \{1,\ldots,p\}$ and a set of undirected edges
$E \subseteq \{ \, \{i,j\} :  i,j =  1,\ldots,p,\ i < j \}$. 
Let $\Ss_p$ denote the set of all symmetric $p \times p$ matrices and 
$\Sp$ the set of all positive definite, symmetric $p \times p$ matrices.
For any graph $G = (V,E)$, let $\Ss^+_p(G)$ be the set of matrices $A \in \Sp$ with zero entries at off-diagonal positions specified by $G$, i.~e., $A_{i,j} = 0$ for all $i,j =  1,\ldots,p$, $i \neq j$, with $\{i,j\} \notin E$. 
We call any set of $p$-dimensional probability measures with the common property that they possess a concentration matrix $K \in \Ss^+_p(G)$ a \emph{covariance selection model} induced by $G$. We call a covariance selection model consisting of all regular $p$-variate Gaussian distributions a \emph{Gaussian graphical model} and denote it by $N_p(G)$, i.~e., 
$N_p(G) = \{ N_p(\bmu,\Sigma) : \bmu \in \R^p, \Sigma^{-1} \in \Ss^+_p(G) \}$. 
The maximum-likelihood estimator $\hat\Sigma_G$ of $\Sigma$ within the parametric family $N_p(G)$ is given as the solution to 
\begin{equation} \label{eq:op}
  \hSigma_G =  \argmin_{\Sigma^{-1} \in \Sp(G)}
	 \left\{ \log \det \Sigma  
									+ \frac{1}{n} \sum_{i=1}^n  
									\trace\left( \hat\Sigma_n \, \Sigma^{-1} \right)
	 \right\}. 
\end{equation}
This optimization problem leads to the estimation equations
\[
	\begin{cases}
	 \  \left[\hSigma_G\right]_{i,j} = \left[\hSigma_n\right]_{i,j} 		&	\qquad  \mbox{ for }  \{i,j\} \in E \  \mbox{ or }  \ i = j, \\[1.0ex]
	 \  \left[\hSigma_G^{-1}\right]_{i,j} = 0               & \qquad  \mbox{ for }  \{i,j\} \notin E \ \mbox{ and } \ i \neq j. \\
	\end{cases}
\]
The solution $\hSigma_G$ depends on the data only through the sample covariance matrix $\hat\Sigma_n$. This approach is due to \citet{Dempster1972}, and the optimization problem (\ref{eq:op}) has since been thoroughly studied \citep[e.~g.][]{Speed1986}. Algorithms to compute $\hSigma_G$ for arbitrary graphs $G$ can be found, e.~g., in \citet[][Chapter 5]{Lauritzen1996} or \citet[][Chapter 17]{Hastie2009}. For decomposable graphs $G$, there is also an explicit solution, i.~e., $\hSigma_G$ can be computed in a finite number of steps. For details see also \citet[][Chapter 5]{Lauritzen1996}. In R \citep{R}, these algorithms are implemented in the function \verb|fitConGraph| in the package ggm \citep{ggm}. 
With a graph-constrained\footnote{i.~e., it obeys the zero-pattern in the inverse induced by $G$} estimate $\hSigma_G$ for the covariance matrix available, the remaining non-zero partial correlations are computed from $\hSigma_G$ as unconstrained partial correlations are computed from $\hSigma_n$ by virtue of (\ref{eq:partial}).

\paragraph{Task (T2)}
Within the parametric framework described above, the likelihood-ratio test for testing $G$ against the full model is given by the test statistic
\begin{equation}
	D^\Sigma_n(G) = n \left(\log\det\hSigma_{G} - \log\det\hSigma_n\right),
\label{eq:deviance}
\end{equation}
which, under the null hypothesis that $G$ is the true graph, converges to a $\chi_q^2$ distribution as $n \to \infty$, where $q$ is the number of missing edges in $G$. The quantity $D_n^\Sigma$ is also called \emph{deviance} and this likelihood-ratio test hence \emph{deviance test}. 
It simultaneously tests the absence of all edges not in $G$ avoiding any multiple-testing problems. The deviance is also returned by \verb|fitConGraph|.

\paragraph{Task (T1)}
While statistical theory provides rather precise and unambiguous solutions to tasks (T3) and (T2), this is not the case for (T1), which is already reflected by the phrasing of \emph{finding an appropriate graph} rather than \emph{finding the best fitting graph}. 
Deciding on an appropriate graph may also be influenced by interpretability aspects and relevant domain knowledge. 
%
%
A multitude of approaches exist. A basic idea, also initiated by \citet{Dempster1972} is the iterative application of the deviance test. For instance, one starts with the full graph, then removes one or several edges (with small absolute partial correlations), and keeps the new candidate graph if the deviance test accepts it. This may be iterated until no further edge removal leads to an accepted graph. The opposite search direction is also possible: One starts with the empty graph and successively adds edges until a graph is obtained which is accepted by the deviance test. For further reading see the textbooks by \citet{Whittaker1990} or \citet{Edwards2000}. Elaborate model search strategies have been proposed \citep[e.~g.][]{Edwards1985, Drton2008}.

Many other model selection approaches use $L_1$-regularization and are aimed at finding sparse graphs in high-dimensional settings. For instance, \citet{Meinshausen2006} propose a node-wise LASSO-regression.
\citet{Yuan2007} and \citet{Friedman2008} add an $L_1$-penalty for $K = \Sigma^{-1}$ to the optimization problem (\ref{eq:op}). Various algorithms have since been proposed for an efficient computation of such high-dimensional optimization problems \citep[e.g.][]{Yuan2010, Cai2011, Sun2013}.
Within this framework, the regularization parameter must be chosen, usually be means of cross validation. \citet{Liu2017} particularly address the latter issue.

\subsection{Robustness}
\label{sec:rob}

Robustness in general terms is the property of a statistical method to yield sensible results if its assumptions are violated. In a more specific sense, it means insensitivity to outliers. Starting with the pioneering work of \citet{Huber1964} and \citet{Hampel1971,Hampel1974}, robust statistics has evolved into a large research area, see, e.~g., the textbooks by \citet{Huber2009} or \citet{Maronna2019}.
\begin{figure}[t]
\caption{The non-robust sample covariance matrix (solid line) and the robust $t_3$ M-estimate (dashed line). The data on the left-hand and right-hand panel differ only by one point.}
\includegraphics[width=0.49\linewidth]{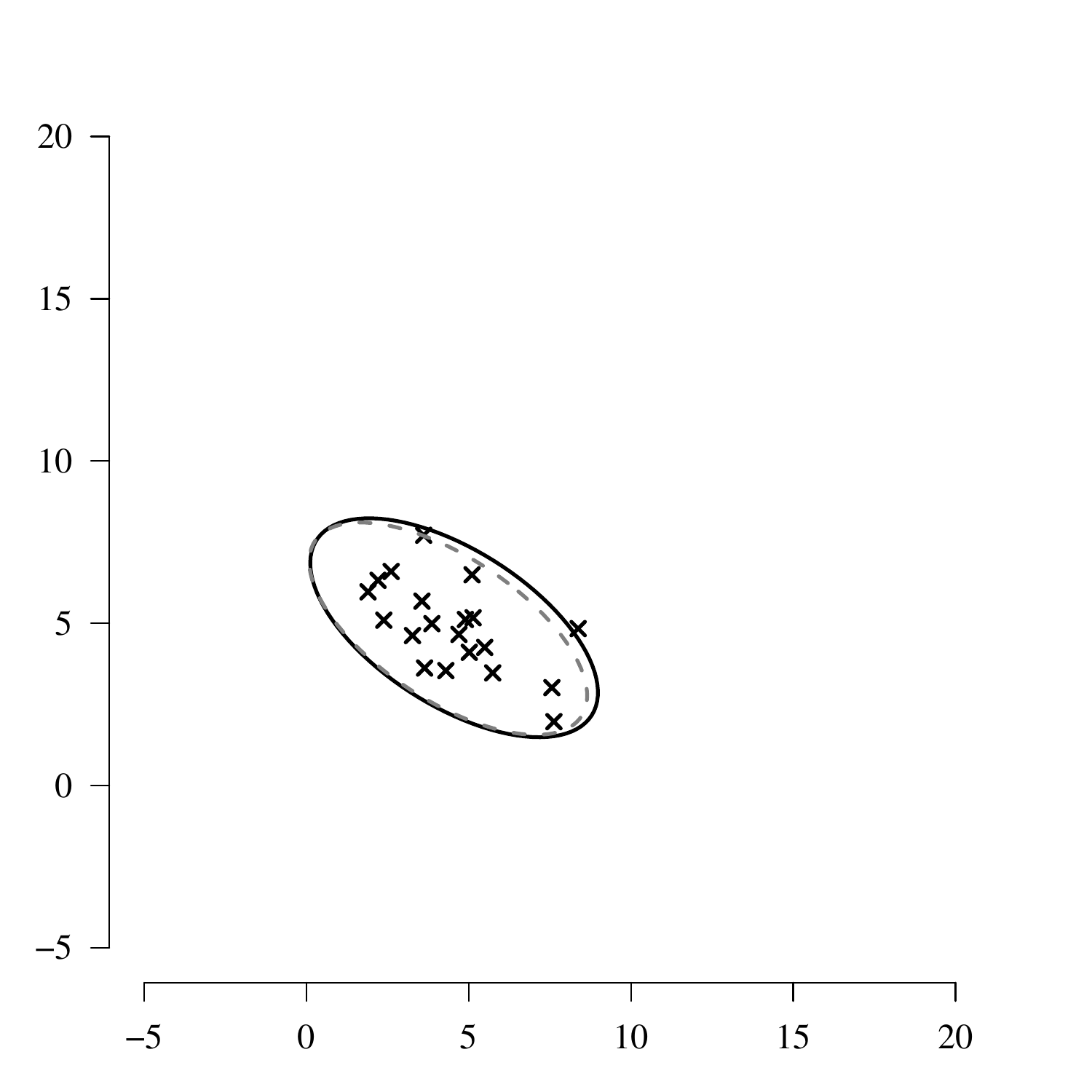}
\includegraphics[width=0.49\linewidth]{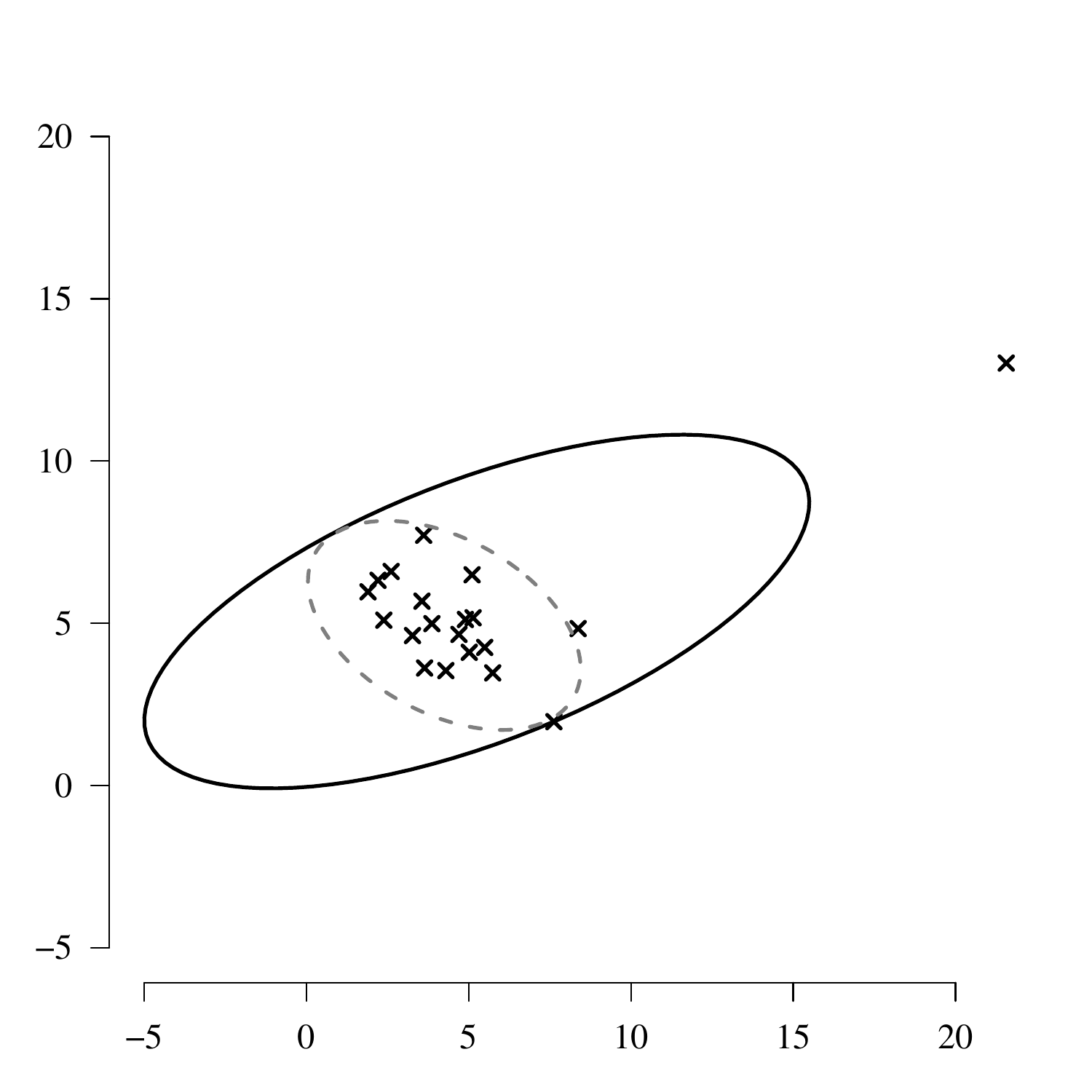}
\label{fig:1}
\end{figure}
For our purposes, the important fact to note is that the sample covariance matrix $\hSigma_n$ is not robust. In Figure \ref{fig:1}, the left-hand panel shows a small  data set of 20 two-dimensional observations. The black ellipse visualizes the sample covariance matrix, i.~e., the 95\% probability ellipse of the thus fitted normal model. In the right-hand panel, one single observation has been moved from the center to the upper right corner. The covariance estimate has tremendously changed, suggesting even a positive rather than a negative correlation. The dashed curve, in contrast, represents an alternative, robust estimator of multivariate scatter (a $t_3$ M-estimator, see below) and is little altered by the outlier.

With the sample covariance matrix $\hSigma_n$ being the main ingredient of essentially all graphical modeling tasks, they all inherit its lack of robustness.
%
The good news is: robust alternatives exist.
The work on robust multivariate location and scatter estimation as been originated by \citet{Maronna1976}, who developed Huber's M-estimation approach for the multivariate setting. 
Since then, many proposals have been made.\footnote{with numerous contributions by David Tyler \citep[e.~g.][]{Tyler1987, Kent1996}}. 
Here we consider only one rather simple and easy-to-compute robust scatter estimator, the $t_\nu$ M-estimator, which is also already mentioned in Maronna's paper. This is an M-estimator, whose loss function stems from the maximum-likelihood estimator within the elliptical $t_\nu$-model. The parameter $\nu$ is referred to as the \emph{degrees of freedom} and may be any positive real number. The smaller the $\nu$, the heavier-tailed the $t_\nu$ distribution and, consequently, the more outlier-resistant the corresponding M-estimator. The parameter $\nu$ is usually not inferred from the data, but selected by the data analyst. A common choice is $\nu = 3$. This is not extremely heavy-tailed, second moments are finite (and hence the covariance matrix is properly defined), but it is sufficiently heavy-tailed to yield a strongly outlier-resistant estimator.

The $t_\nu$ M-estimator of scatter $\hat{S}_n$, along with the corresponding estimate of location $\hat\bmu_n$, is defined as the solution to the optimization problem
\begin{equation} \label{eq:M1}
	(\hat{\bmu}_n, \hS_n) = \argmin_{\bmu \in \R^p, S \in \Sp} 
	\left[ \sum\nolimits_{i=1}^n \rho_{\nu,p}\left\{ (\bx_i - \bmu)^\top S^{-1} (\bx_i - \bmu)  \right\} + n \log \det S \right]
\end{equation}
with $\rho_{\nu,p}(x) = (\nu + p) \log(1 + x/\nu)$. This yields the estimation equations
\[
\begin{cases}
	\displaystyle
 \  0 = \sum\nolimits_{i=1}^n  \psi_{\nu,p}(\hat{r}_i)(\bx_i - \hat\bmu_n) ,  \\[8pt] 
    \displaystyle
 \  \hS_n = n^{-1} \sum\nolimits_{i=1}^n \psi_{\nu,p}(\hat{r}_i) (\bx_i-\hat\bmu_n)(\bx_i-\hat\bmu_n)^\top, \\
\end{cases}
\]
where $\psi_{\nu,p}(x) = \rho'_{\nu,p}(x) = (\nu + p)/(\nu + x)$ and $\hat{r}_i = (\bx_i-\hat\bmu_n)^\top \hS_n^{-1}(\bx_i-\hat\bmu_n)$.
The $t_\nu$ M-estimator can be computed, e.~g., by a fixed-point algorithm. Implementations in R can be found, e.~g., in the functions \verb|cov.trob| from the package MASS \citep{MASS}, \verb|tM| from the package ICS \citep{Nordhausen2008}, and \verb|MVTMLE| from the package fastM \citep{Dumbgen2016, fastM}. The latter uses a partial Newton-Raphson algorithm.

With robust alternatives $\hS_n$ being available, an intuitive path to a robust analysis is the plug-in approach: First solving (\ref{eq:M1}) and then plugging the thus obtained estimate $\hS_n$ instead of $\hSigma_n$ into (\ref{eq:op})  to obtain a graph-constrained robust estimate $\hS_G$ and the corresponding robust (pseudo-)deviance
\begin{equation} \label{eq:pseudo}
		D^S_n(G) = n \left(\log\det\hat{S}_G - \log\det\hat{S}_n \right).
\end{equation}
This is accomplished in the package robFitConGraph by the function of the same name. 
Plug-in robustifications for the $L_1$-regularization methods have equally been proposed \citep{Ollerer2015, Tarr2016}. Alternatively, \citet{Finegold2011} regularize the elliptical $t_\nu$-log-density. 

Before providing further details on the function \verb|robFitConGraph| and the underlying theory in Section \ref{sec:the}, its use shall be demonstrated at a data
example on music performance anxiety. 

%
%

%
%
%
%
%
%
\section{A Case Study: Music Performance Anxiety}
\label{sec:mpa}

The fear about one's ability to perform a specific task, such as giving a presentation or sitting an exam, affects almost everyone. Pressure can be particularly high in certain professions where performing in front of others is an integral part of day-to-day life. While some levels of stress and anxiety are normal and actually help us to achieve optimal performance, severe levels of stress and anxiety are debilitating and can develop into a disorder. Professional musicians are often exposed to extreme pressure where maintaining top quality performances are not just essential to keeping their job, but to progress in their careers. Music performance anxiety (MPA) can be understood as a continuum ranging from low to high anxiety levels. The latter poses a serious problem to the profession and is the subject of ongoing clinical research, see, e.~g. \citet{Fernholz2019} for a recent review.

MPA is often considered to be a form of social anxiety which, loosely speaking, is the overwhelming fear of social situations 
\citep[e.~g.][]{Kenny2011, Cox1993, Nicholson2015, Dobos2019}. This is underlined by the fact that the Diagnostic and Statistical Manual of Mental Disorders \citep[DSM-5;][]{DSM5} now acknowledges evidence of individuals suffering exclusively from performance anxiety as a distinct sub-type of social anxiety disorder.
However, some researchers and clinicians have questioned this description, as MPA is a complex phenomenon caused by the interaction of many different factors, and the fear of social judgment is not necessarily always the main problem.

\citet*{Wiedemann2021} analyzed a data set consisting of eight numerical variables measured at $n = 82$ students at German music colleges: music performance anxiety (MPA), agoraphobia (AG), generalized anxiety disorder (GAD), panic disorder (PD), separation anxiety disorder (SEP), specific phobia (SP), social anxiety disorder (SAD) as well as illness anxiety disorder (ILL). Each variable is a summary score from a self-assessment inventory with Likert-scale items. A higher value signifies a higher severity of the condition. 
MPA was assessed using the German version of the Kenny Music Performance Anxiety Inventory \citep[K-MPAI;][]{Kenny2009} 
translated by \citet{Spahn2016}.
All other anxieties were assessed using the German 
translation of the disorder-specific anxiety measures \citep{Beesdo2012, Lebeau2012}
for the dimensional anxiety scales of the DSM-5. 
The data set is also included in the package robFitConGraph as \verb|anxieties|.
\begin{figure}[t]
\caption{Pairwise scatter plots of the anxieties data}
\includegraphics[width=0.9\linewidth]{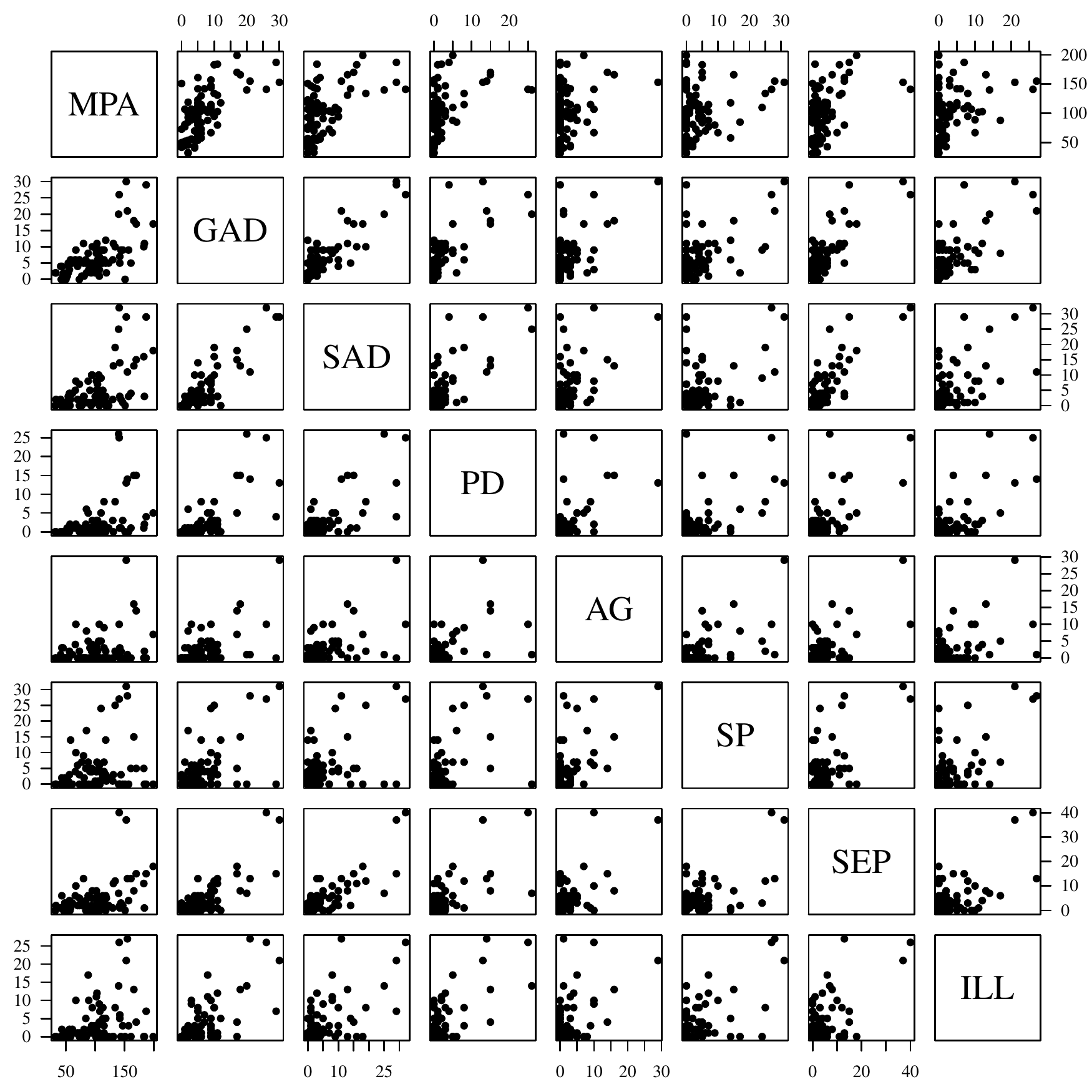}
\label{fig:2}
\end{figure}

Figure \ref{fig:2} shows the pairwise scatter plots of the data set. While most participants score low on most anxiety scales -- as we would expect and hope --, there are a few very high values in all variables. The normality assumption can
be seen to be violated in a similar manner as in Figure \ref{fig:1} with the same implications for any sample-covariance-based analysis. 
For instance, removing the outlier in the variable AG reduces the Pearson correlation between AG and SEP from 0.612 to 0.407. Computing the correlation coefficient from a $t_3$ M-estimator, we obtain the value 0.410, which reduces to 0.350 when removing said outlier. A robust analysis is highly recommended for this data. All results reported in the following are based on a $t_3$ M-estimator of scatter. 
It yields the correlation coefficients given in Table \ref{tab:1}. They can 
also be obtained by the function \verb|robFitConGraph| by supplying the full model as adjacency matrix
\begin{verbatim}
> library(robFitConGraph)
> data(anxieties)
> p <- ncol(anxieties)
> Shat <- robFitConGraph(X = anxieties,
+     amat = matrix(1, ncol = p, nrow = p),
+     df = 3)$Shat
> round(cov2cor(Shat), d = 2)
\end{verbatim}
\begin{table}[h]
\caption{Pairwise correlations computed from $t_3$-M-estimate}
\label{tab:1}   
\begin{tabular}{p{0.08\linewidth}p{0.08\linewidth}p{0.08\linewidth}
								p{0.08\linewidth}p{0.08\linewidth}p{0.08\linewidth}p{0.08\linewidth}p{0.08\linewidth}p{0.08\linewidth}}  
\hline\noalign{\smallskip}
           &      MPA &      GAD &      SAD &       PD &       AG &       SP &      SEP &      ILL \\ 
\noalign{\smallskip}
 MPA       &     $\, \cdot \,$ &     0.62 &     0.37 &     0.43 &     0.17 &     0.26 &     0.38 &     0.32 \\ 
 GAD       &     0.62 &     $\, \cdot \,$ &     0.66 &     0.65 &     0.40 &     0.35 &     0.60 &     0.47 \\ 
 SAD       &     0.37 &     0.66 &     $\, \cdot \,$ &     0.50 &     0.51 &     0.36 &     0.64 &     0.42 \\ 
 PD        &     0.43 &     0.65 &     0.50 &     $\, \cdot \,$ &     0.51 &     0.44 &     0.48 &     0.46 \\ 
 AG        &     0.17 &     0.40 &     0.51 &     0.51 &     $\, \cdot \,$ &     0.49 &     0.41 &     0.36 \\ 
 SP        &     0.26 &     0.35 &     0.36 &     0.44 &     0.49 &     $\, \cdot \,$ &     0.37 &     0.34 \\ 
 SEP       &     0.38 &     0.60 &     0.64 &     0.48 &     0.41 &     0.37 &     $\, \cdot \,$ &     0.29 \\ 
 ILL       &     0.32 &     0.47 &     0.42 &     0.46 &     0.36 &     0.34 &     0.29 &     $\, \cdot \,$ \\
\noalign{\smallskip}\hline\noalign{\smallskip}
\end{tabular}
\end{table}
%
%
%

All variables are positively correlated. This is neither surprising nor uncommon for multivariate data.
The corresponding partial correlations, as given in Table \ref{tab:2}, shed further light on the dependence structure of the variables. 
\begin{verbatim}
> Phat <- -cov2cor(solve(Shat))
> diag(Phat) <- 1
> round(Phat,d = 2)
\end{verbatim}
\begin{table}[h]
\caption{Pairwise \emph{partial} correlations computed from $t_3$-M-estimate}
\label{tab:2}   
\flushright
\begin{tabular}{p{0.08\linewidth}p{0.08\linewidth}p{0.08\linewidth}
								p{0.08\linewidth}p{0.08\linewidth}p{0.08\linewidth}p{0.08\linewidth}p{0.08\linewidth}p{0.08\linewidth}}  
\hline\noalign{\smallskip}
           &      MPA &      GAD &      SAD &       PD &       AG &       SP &      SEP &      ILL \\ 
\noalign{\smallskip}
MPA       &     $\ \cdot \,$ &     0.44 &    -0.05 &     0.05 &    -0.14 &     0.08 &     0.05 &     0.05 \\ 
GAD       &     0.44 &     $\ \cdot \,$ &     0.33 &     0.34 &    -0.02 &    -0.06 &     0.19 &     0.15 \\ 
SAD       &    -0.05 &     0.33 &     $\ \cdot \,$ &    -0.05 &     0.25 &     0.00 &     0.37 &     0.13 \\ 
PD        &     0.05 &     0.34 &    -0.05 &     $\ \cdot \,$ &     0.24 &     0.14 &     0.09 &     0.16 \\ 
AG        &    -0.14 &    -0.02 &     0.25 &     0.24 &     $\ \cdot \,$ &     0.30 &     0.04 &     0.07 \\ 
SP        &     0.08 &    -0.06 &     0.00 &     0.14 &     0.30 &     $\ \cdot \,$ &     0.11 &     0.12 \\ 
SEP       &     0.05 &     0.19 &     0.37 &     0.09 &     0.04 &     0.11 &     $\ \cdot \,$ &    -0.12 \\ 
ILL       &     0.05 &     0.15 &     0.13 &     0.16 &     0.07 &     0.12 &    -0.12 &     $\ \cdot \,$ \\ 
\noalign{\smallskip}\hline\noalign{\smallskip}
\end{tabular}
\end{table}
%
%
%
Many of the partial correlations are near zero, suggesting conditional independences, i.~e., their association may be fully mediated by other variables in the data set.

\subsection{Inferential analysis: MPA and social anxiety}

One hypothesis examined by \citet{Wiedemann2021} is whether MPA is primarily related to a social anxiety disorder (SAD). 
Based on the positive correlation of $0.37$, a short-sighted analysis may deduce a strong connection between MPA and SAD, which in light of other, even larger correlations, is right away questionable. As we will see, the data in fact carries no evidence for a particularly strong connection between MPA and SAD. 

For that purpose one tests the hypothesis that MPA and SAD are conditionally independent given all remaining six variables, i.~e., testing a graphical model with only one missing edge between MPA and SAD by means a robust pseudo-deviance test. If this test rejects, there is evidence for a strong link between MPA and SAD. Their dependence can then not be fully explained by their associations with other anxiety types. However, this hypothesis is \emph{not} rejected (a p-value of 0.66), and we find no such evidence in the data. The p-value is returned by the function \verb|robFitConGraph| via the named list element \verb|pval|, see also the next code chunk below.

\begin{figure}[t]
\caption{The graph encoding the hypothesis MPA is conditionally independent of all other variables given GAD (left) and the corresponding adjacency matrix (right).}
\smallskip

\begin{minipage}[c]{0.42\linewidth}
\includegraphics[width=\linewidth]{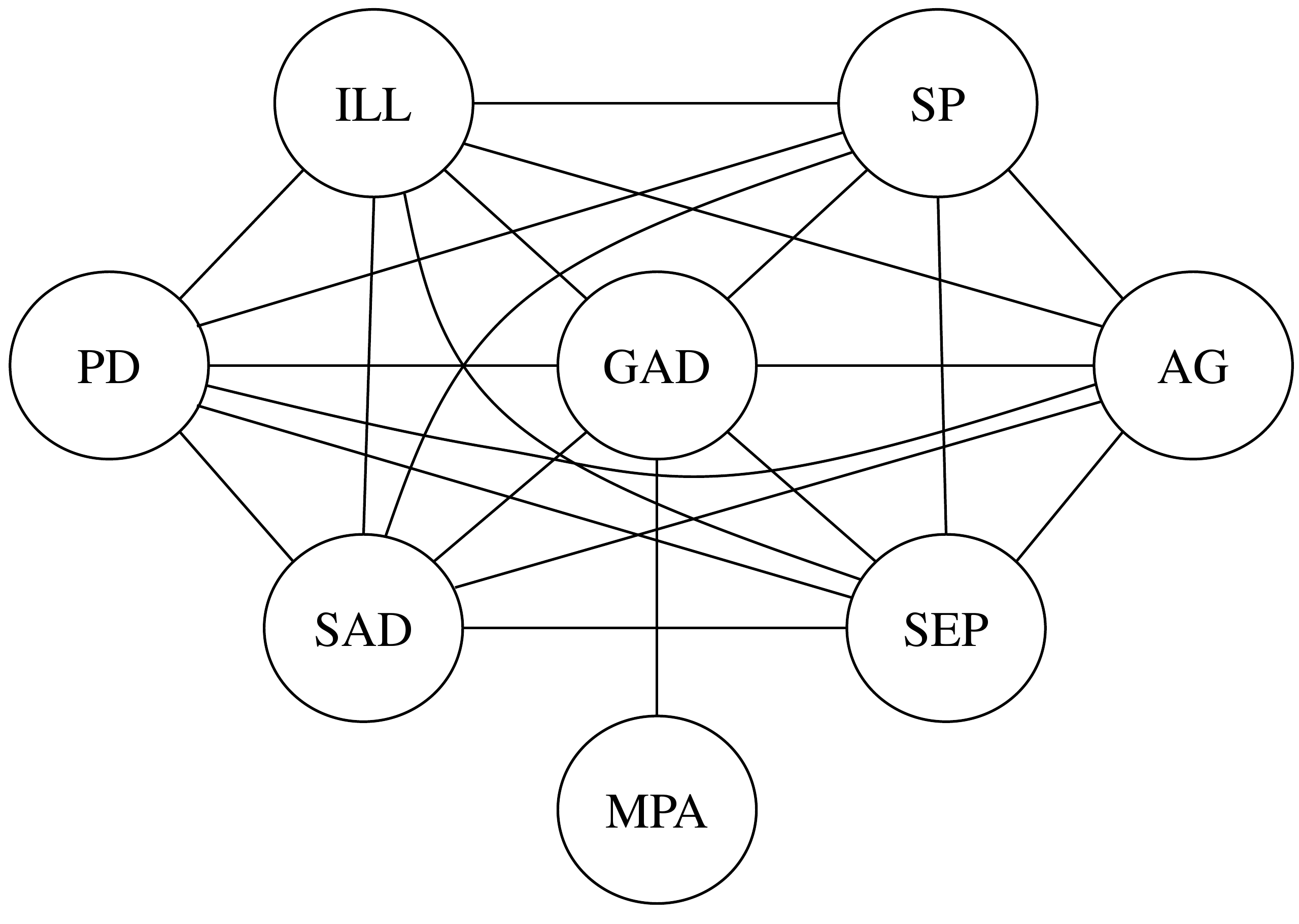}
\end{minipage}
\hfill
\begin{minipage}[c]{0.57\linewidth}
\begin{tabular}{p{0.06\linewidth}p{0.06\linewidth}p{0.06\linewidth}
								p{0.06\linewidth}p{0.06\linewidth}p{0.06\linewidth}p{0.06\linewidth}p{0.06\linewidth}p{0.06\linewidth}}   
\hline\noalign{\smallskip}
           &      MPA &     GAD &      SAD &       PD &       AG &       SP &      SEP &      ILL \\ 
\noalign{\smallskip}
MPA &  1 &  1 &  0 & 0 & 0 & 0 &  0 &  0 \\
GAD &  1 &  1 &  1 & 1 & 1 & 1 &  1 &  1 \\
SAD &  0 &  1 &  1 & 1 & 1 & 1 &  1 &  1 \\
PD &   0 &  1 &  1 & 1 & 1 & 1 &  1 &  1 \\
AG &   0 &  1 &  1 & 1 & 1 & 1 &  1 &  1 \\
SP &   0 &  1 &  1 & 1 & 1 & 1 &  1 &  1 \\
SEP &  0 &  1 &  1 & 1 & 1 & 1 &  1 &  1 \\
ILL &  0 &  1 &  1 & 1 & 1 & 1 &  1 &  1 \\
\noalign{\smallskip}\hline\noalign{\smallskip}
\end{tabular}
\end{minipage}
\label{fig:3}
\end{figure}

One can even further test the stronger hypothesis that MPA is, given GAD only, conditionally independent of all other six specific anxiety scales. This is done by testing the graph in Figure \ref{fig:3}, which can equivalently be expressed by its adjacency matrix on the right-hand side of Figure \ref{fig:3}.

Almost unnoticed we make here use of a non-trivial result, which is the real merit and the real beauty of graphical models: Each absent edge in the graph denotes a conditional independence of two individual variables given the respective remaining six variables. This is indeed equivalent to MPA and (SAD, PD, AG, SP, SEP, ILL) being conditionally independent given GAD only, because, in the graph, GAD separates MPA from (SAD, PD, AG, SP, SEP, ILL). This is known as the equivalence between the local Markov property and the global Markov property. For details see \citet[][Chapter 3]{Lauritzen1996}.

The pseudo-deviance test based on the $t_3$ M-estimator for the graph in Figure \ref{fig:3} is carried out as follows:
\begin{verbatim}
> amat <- matrix(1, ncol = p, nrow = p)
> rownames(amat) <- colnames(anxieties)
> colnames(amat) <- colnames(anxieties)
> amat["MPA", c("SAD", "PD", "AG", "SP", "SEP", "ILL")] <- 0
> amat[c("SAD", "PD", "AG", "SP", "SEP", "ILL"), "MPA"] <- 0
> robFitConGraph(X = anxieties, amat = amat, df = 3)$pval
[1] 0.881159
\end{verbatim}

With a p-value of 0.88, this hypothesis is also \emph{not} rejected. There is no evidence against the null hypothesis of MPA being conditionally independent of the specific anxiety types given generalized anxiety (GAD). 
Accepting the null hypothesis, GAD is fully sufficient for predicting MPA. When already knowing a person's GAD score, i.~e., how anxious the person generally is, additionally knowing any of their specific anxieties scores provides no further information about their MPA score. So this analysis does not support the hypothesis that MPA is foremost related to social anxiety.

The analysis is complemented by testing if MPA and GAD are conditionally independent given the remaining six variables. This hypothesis \emph{is} rejected with a p-value below 0.01, which corresponds to the partial correlation of 0.44 in Table \ref{tab:2}.%
\footnote{For a single missing edge, the deviance test can indeed be expressed in terms of corresponding partial correlation only.}
So there is a strong connection between MPA and GAD, this connection is not mediated by any of the other variables, and GAD explains the connection between MPA and the remaining variables. 
Altogether, GAD is the link between MPA and the other anxieties.

\subsection{Explorative analysis}

\begin{figure}[t]
\caption{A fitting graph with 9 edges based on a $t_3$ scatter estimator with p-value 0.45.}
\flushright\includegraphics[width=0.6\linewidth]{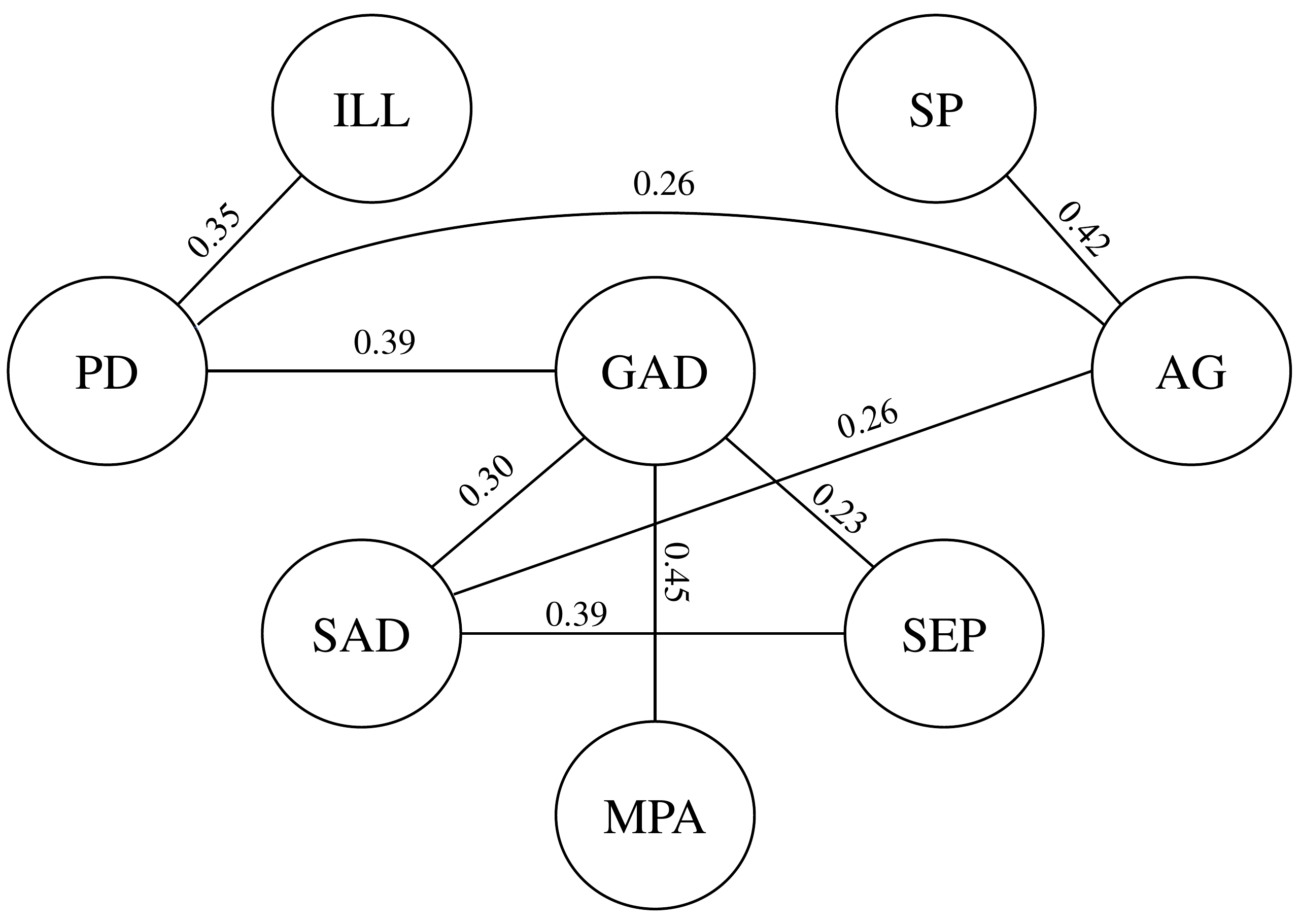}
\label{fig:4}
\end{figure}

Naturally, the question arises which other edges may be removed. For that purpose we remove all edges which have absolute partial correlation below 0.15 (Table~\ref{tab:2}), corresponding to an individual p-value above 0.2. The resulting graph is depicted in Figure \ref{fig:4}, which is not rejected by the pseudo-deviance test with a p-value of 0.45. 
The partial correlations along the edges in Figure \ref{fig:4} are fitted under the graph, i.~e., they are estimated taking the graph structure into account. They are different from the partial correlations given in Table \ref{tab:2}.
\begin{verbatim}
> amat <- abs(Phat) > 0.15
> Shat_G <- robFitConGraph(X = anxieties, 
+     amat = amat, df = 3)$Shat
> Phat_G <- -cov2cor(solve(Shat_G))
> diag(Phat_G) <- 1
> round(Phat_G, d = 2)									
\end{verbatim}
The information learned from such a fitted graph is, e.~g., that GAD is central within this set of variables: It has many edges to other vertices, it has much explanatory power about the other variables.
If one were to retain only a single variable (as a very simple dimension reduction approach, say), GAD would be a natural candidate based on this ``vertex degree criterion''.

The fitted graph in Figure~\ref{fig:4} with 9 edges (out of 28 possible) is one parsimonious graph that fits the data. Generally, there is 
no well-defined \emph{most parsimonious}, \emph{best-fitting} graph, as obviously fit and parsimony are contradicting goals. 
However, one may elaborate upon the initial search by checking if other (equally or more) parsimonious graphs may fit as well. 
For instance, increasing the initial partial correlation threshold to 0.17, say, which results in a further removal of the edge ILL--PD, leads to a p-value of 0.01 and hence should be rejected. Alternatively, removing the ``next smallest edge'' from the candidate graph in Figure~\ref{fig:4}, which is GAD--SEP, results in a p-value of 0.17, which is still acceptable. 

Despite using deviance-test p-values as decision criterion whether to accept a graph or not, this analysis is purely explorative. We have not \emph{tested} the graph of Figure~\ref{fig:4}. Hypotheses to be tested must be formed a-priori. Testing for a graph which is the result of model selection procedure is as prohibitive as testing if the observed sample mean is the population mean.

So far, we have used \verb|robFitConGraph| with \verb|df = 3|, which is also the default setting in case \verb|df| is not specified.
Generally, the results are not very sensitive to variations in $\nu$. 
A smaller value of $\nu$ downweights outliers more strongly.
In the present example, taking $\nu = 1$, we find the hypothesis of Figure~\ref{fig:3} equally accepted with a p-value of 0.58. The explorative graph in Figure~\ref{fig:4} is accepted with a p-value of 0.26.

\subsection{The classical analysis}

\begin{figure}[t]
\caption{A fitting graph with 10 edges based on the sample covariance matrix with p-value 0.51.}
\flushright\includegraphics[width=0.6\linewidth]{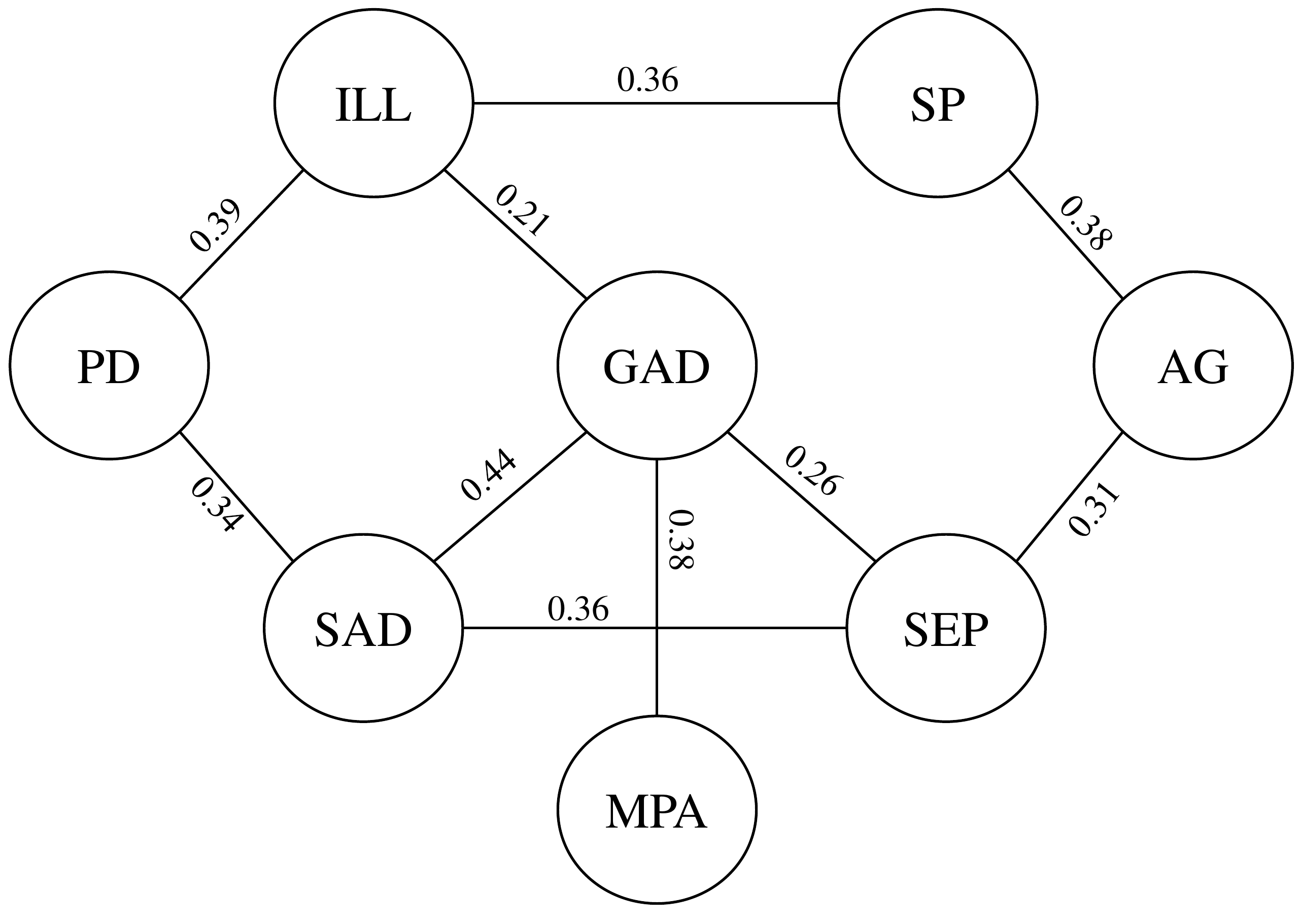}
\label{fig:5}
\end{figure}

After having cautioned its use at the beginning, we close this section by remarking that a classical sample-covariance-based analysis leads to qualitatively the same findings.
The classical deviance test gives a p-value of 0.63 for the hypothesized graph of Figure~\ref{fig:3}. 
An explorative analysis analogous to the one above leads to the graph in Figure~\ref{fig:5} with a p-value of 0.51, which is also communicated by \citet{Wiedemann2021}. 
The graph is different, but it equally shows the centrality of GAD as well as the marginality of MPA within the set of variables. 
It is generally advisable to try varying parameter settings and different methods, robust and non-robust, in any statistical analysis. 
%
Conclusions unanimously obtained by several methods may be considered even more trustworthy. 

%
%
%
%
%
%
%
%
\section{Background and theory}
\label{sec:the}

The centerpiece of the package robFitConGraph is the function by the same name. For a given data set $\dsX_n$ and a given graph $G$, it simultaneously provides a robust graph-constrained matrix $\hat{S}_G$ and the p-value of the corresponding pseudo-deviance test. But so far, the actual worth of the function \verb|robFitConGraph| has not become apparent: It appears, its plug-in functionality is equally achieved by a combination of, say, \verb|cov.trob| and \verb|fitConGraph|, as the latter takes the sample covariance matrix as input. However, there are two good reasons for \verb|robFitConGraph|, which both require a slightly deeper foray into statistical theory.

\begin{enumerate}[(1)]
\item
The limit distribution of the pseudo-deviance $D^S_n(G)$ does not follow strictly a $\chi^2_q$-distribution under the graph $G$, but a $\chi^2_q$-variate multiplied by a constant $\sigma_1$. This constant is required to obtain p-values and depends on the data dimension $p$ and the degrees of freedom $\nu$ of the estimator. It is computed by the function \verb|find_sigma1|.
\item
The plug-in estimator $\hat{S}_G$ is just one approach to a graph-constrained robust scatter estimator. The function \verb|robFitConGraph| also provides an alternative approach, referred to as the \emph{direct estimator} and denoted by $\tilde{S}_G$ below.
\end{enumerate}
Another good reason is speed. Being implemented in C++, \verb|robFitConGraph| is considerably faster than the combination of \verb|cov.trob| and \verb|fitConGraph|. A runtime comparison is given by \citet{Watt2019}.

\subsection{The Constant $\sigma_1$}

Recall the class of all $p$-dimensional, continuous, elliptical distributions, i.~e., distributions possessing a $p$-dimensional Lebesgue density $f$ of the form
\begin{equation} \label{eq:density}
	f(\bx) = \det(S)^{-\frac{1}{2}} g\big\{(\bx-\bmu)^\top S^{-1} (\bx-\bmu)\big\}
\end{equation}
for some $\bmu \in \R$, $S \in \Sp$ and $g:[0,\infty) \to [0,\infty)$, such that $f$ integrates to 1.
Let $E_p(\bmu,S,g)$ denote the distribution described by (\ref{eq:density}). The univariate function $g$ is called the \emph{elliptical generator} and $S$ the \emph{scatter} or \emph{shape matrix} of $E_p(\bmu,S,g)$. It is proportional to the covariance matrix if $E_p(\bmu,S,g)$ has second moments.
In the present paper, we only consider two examples of elliptical distributions: the normal distribution with
\[
	g_N(x) = \frac{1}{\sqrt{2\pi}} \exp\left( - \frac{x}{2}\right), \qquad 0 \le x,
\]
and the elliptical $t_\nu$-distribution with
\[
	g_{\nu,p}(x) = c_{\nu,p} \left( 1 + \frac{x}{\nu} \right)^{-(\nu+p)/2}, \qquad 0 \le x.
\]
where the normalizing constant $c_{\nu,p}$ is given in the Technical Appendix at the end of the chapter. 
Assume the independent observations $\dsX_n = (\bx_1, \ldots \bx_n)^\top$ to stem from an elliptical distribution $E_p(\bmu,S,g)$ and let
$\hV_n$ be an \emph{arbitrary} $p \times p$ scatter matrix estimator fulfilling two fairly natural conditions:
\begin{enumerate}[(1)]
	\item $\hV_n$ is affine equivalent, i.~e., $\hV_n(\dsX_n A^\top + \bOne_n \bb^\top ) =  A \hS_n(\dsX_n) A^\top$
  for any $\bb \in \R^p$ and full rank $A \in \R^{p \times p}$, where $\bOne_n$ is the $n$-vector consisting of ones. 
	\item $\hV_n$ is $\sqrt{n}$-consistent and asymptotically normal at $E_p(\bmu,S,g)$, i.~e., there exist matrices $V \in \Ss_p$ and
	and $W \in \Ss_{p^2}$ such that
	\[
			\sqrt{n} \vv \{ \hV_n(\dsX_n) - V \} \ \to \ N_{p^2}\! \left( \bZero, \ W \right) 
	\]
	in distribution as $n \to \infty$.
\end{enumerate}
\citet{Tyler1982} showed that under these two conditions, $V$ and $W$ take on specific forms:
$V = \eta S$ for some $\eta > 0$, and
\begin{equation} \label{eq:Mp}
   W = 2 \eta^2 \sigma_1 \mM_p (S\otimes S)\ + \ \eta^2 \sigma_2 \vv(S) \vv(S)^\top,
\end{equation}
where $\otimes$ is the Kronecker product,  
$\sigma_1 \ge 0$ and $\sigma_2 \ge - 2\sigma_1/p$ 
are scalar constants independent of $\bmu$ and $S$, and
$\mM_p$ is a fixed $p^2 \times p^2$ matrix defined in the Technical Appendix.
%
The latter formula greatly simplifies the asymptotic efficiency comparison of any two affine equivariant scatter estimators at elliptical distributions. 
In case of $\hV_n$ being an elliptical M-estimator, with a general loss function $\rho$ instead of $\rho_{\nu,p}$ in (\ref{eq:M1}), \citet{Tyler1983} gives the following expressions for the scalars $\eta$, $\sigma_1$ and $\sigma_2$:
Letting $\psi(x) = \rho'(x)$, $\phi(x) = x \psi(x)$, and 
$R = (\bX - \bmu)^\top S^{-1} (\bX-\bmu)$ for $\bX \sim E_p(\bmu,S,g)$, the scalar $\eta$ is the solution to 
   $\E\{\phi(R/\eta)\} = p$.
Letting further 
\[
	\gamma_1 = \frac{\E\{\phi^2(R/\eta)\}}{p(p+2)}, 
	\qquad
	\gamma_2 = \frac{1}{p} \E\left\{\frac{R}{\eta}\phi'\left(\frac{R}{\eta}\right)\right\},
\]
the scalars $\sigma_1$ and $\sigma_2$ are
\[
	\sigma_1  =  \frac{(p+2)^2 \gamma_1}{(2\gamma_2 + p)^2},  \qquad 
	\sigma_2  =  \gamma_2^{-2}\left[ \gamma_1 -1 - \frac{2\gamma_1(\gamma_2 -1)\{p+(p+4)\gamma_2\}}{(2\gamma_2 + p)^2} \right].
\] 
\citet{Tyler1983} further showed that the asymptotic variance of any scale-invariant, continuously differentiable function $h$ of $\hV_n$ only depends $\sigma_1$ and not on $\sigma_2$. A \emph{scale-invariant} function $h:\Ss_p \to \R$ satisfies $h(\alpha V) = h(V)$ for any $\alpha > 0$. The pseudo-deviance test statistic (\ref{eq:pseudo}) is such a scale-invariant function of the scatter estimator $\hS_n$.
Dependence is an inherently scale-free concept. So this equally applies to any aspect of multivariate scatter that quantifies dependence in one way or another, may it be correlations, partial correlations, canonical correlations, principal components, etc.

In the package robFitConGraph, the functions \verb|find_eta| and \verb|find_sigma1| compute the scalars $\eta$ and $\sigma_1$, respectively, for $t_{\nu_1}$ M-estimators in case the data stem from a normal or an elliptical $t_{\nu_2}$ distribution. Note that the degrees of freedom $\nu_1$ of the estimator loss function $\rho_{\nu_1,p}$ and the degrees of freedom of the population distribution $\nu_2$ can be generally different. If they coincide, the $t_{\nu_1}$ M-estimator is a maximum-likelihood estimator.

The value of $\nu_1$ is specified by \verb|df_est| and $\nu_2$ by \verb|df_data|. For both, \verb|Inf| is allowed, which corresponds to the sample covariance matrix and the normal distribution, respectively. For \verb|find_sigma1|, the input \verb|df_est = 0| is also allowed. This corresponds to Tyler's distribution-free M-estimator of scatter \citep{Tyler1987}. In this case $\sigma_1 = 1 + 2/p$ regardless of the elliptical population distribution. There is no $t_{0}$ distribution, hence \verb|df_data = 0| is not allowed. 

The value $\sigma_1$ can be given directly to \verb|robFitConGraph| via the optional argument \verb|sigma1|. If none is provided, 
\verb|robFitConGraph| calls \verb|find_sigma1| with \verb|df_data = Inf| (i.~e., assuming Gaussian data) and its argument \verb|df| being passed on as \verb|df_est| to the function \verb|find_sigma1|. The argument \verb|df| of \verb|robFitConGraph| is optional with the default \verb|df = 3|.

\subsection{The Direct vs.\ the Plug-in Estimate}

Instead of solving (\ref{eq:M1}) and (\ref{eq:op}) sequentially, an alternative approach is to directly solve the optimization problem
\begin{equation} \label{eq:M2}
	(\tilde\bmu_{G}, \tilde{S}_{G}) = 
	\argmin_{\bmu \in \R^p, S^{-1} \in \Sp(G)} 
	\left[ 
	   \sum\nolimits_{i=1}^n \rho_{\nu,p}\!\left\{ (\bx_i - \bmu)^\top S^{-1} (\bx_i - \bmu)\right\}  + n \log \det S
	\right],
\end{equation}
with $\rho_{\nu,p}$ as in (\ref{eq:M1}), leading to the estimation equations
\begin{equation} \label{eq:M3}
\begin{cases}
	\displaystyle
 \  0 = \sum\nolimits_{i=1}^n \psi_{\nu,p}(\tilde{r}_{i})(\bx_i - \tilde\bmu_G) ,  \\[8pt] 
  \displaystyle
 \  \left[\tilde{S}_G\right]_{j,k} 
     =  \left[ 
		        n^{-1}\sum\nolimits_{i=1}^n \psi_{\nu,p}(\tilde{r}_i) (\bx_i - \tilde\bmu_G)(\bx_i - \tilde\bmu_G)^\top 
				\right]_{j,k},
 			&	  \mbox{for } \{j,k\} \in E \  \mbox{ or }  \ j = k, \\[10pt]
 	\displaystyle		
 \  \left[ \tilde{S}_G^{-1} \right]_{j,k}  = 0,     
 			&    \mbox{for } \{j,k\} \notin E \ \mbox{ and } \ j \neq k,
 \end{cases}
\end{equation}
where $\tilde{r}_i = (\bx_i-\tilde\bmu_G)^\top \tilde{S}_n^{-1}(\bx_i - \tilde\bmu_G)$ and, as before, $\psi_{\nu,p}(x) = \rho'_{\nu,p}(x)$.

The estimator $\tilde{S}_G$ shall be called the \emph{direct estimator}, which is short for 
\emph{direct graph-constrained $t_{\nu}$ M-estimator}, and is an alternative to the plug-in graph-constrained $t_{\nu}$ M-estimator.
Using the function \verb|robFitConGraph|, the direct estimator is invoked by setting the option \verb|direct = TRUE| or \verb|plug_in = FALSE|. In case of conflicting specifications, \verb|plug_in| has priority, and a message will be displayed.
Contrary to the plug-in estimator $\hS_G$, the direct estimator $\tilde{S}_G$ is not a function of the corresponding unconstrained estimate $\hS_n$ alone. 

One main theoretical result of \citet{Vogel2014} is the asymptotic equivalence of $\hS_G$ and $\tilde{S}_G$ under elliptical population distributions. Hence the limiting distribution of the pseudo-deviance and the constant $\sigma_1$ are the same in both cases.
It may be argued that this asymptotic equivalence result favors the plug-in estimator: Considering the elliptical distribution (\ref{eq:density})
with the fixed generator $g_{\nu,p}$, one defines the elliptical-$t_{\nu}$ graphical model $E_p(g_{\nu,p}, G)$ analogously to the Gaussian graphical model $N_p(G)$. The direct graphical $t_\nu$ M-estimator $(\tilde{\bmu}_G,\tilde{S}_G)$ with the corresponding loss function $\rho_{\nu,p}$ is then the maximum-likelihood estimator within this parametric family.\footnote{The parameters of interest are $\bmu$ and $S$. The degrees of freedom $\nu$ is held fixed.}
So direct graphical M-estimators generalize maximum-likelihood estimators, which are known to be first-order efficient. 
Plug-in estimators are popular with practitioners as they are easily applied, 
fast to compute, and now it turns out that they also possess desirable asymptotic properties. 
Direct graphical M-estimators are generally harder to compute. They may be solved by a double-loop, iteratively reweighted
least-squares algorithm, where the Gaussian model fit is nested into the M-estimation loop.

The other half of the story is that direct graphical M-estimators can be substantially more efficient in small samples as is demonstrated by simulations in \citet{Vogel2014}.

\subsection{Ellipticity vs.\ Normality}

Multivariate data containing outliers may be modeled conceptually in two different ways:
Either by a corrupted Gaussian distribution, i.~e., a few observations are erroneous and stem from a different, outlier-generating distribution, or by a heavy-tailed elliptical distribution, which generates outliers itself.
Using the $t_\nu$ M-estimator 
implicitly suggests the latter viewpoint. However, two issues arise:
\begin{enumerate}[(1)]
\item 
Many data sets exhibit features such as the anxiety data set: It is clearly not normal as it contains outliers, but it is clearly not elliptical either as it is skewed.
Hence we do adopt the viewpoint of corrupted normal data. In the analysis, however, we apply outlier-resistant methods which have been derived from considerations in the elliptical $t_{\nu}$ model. We work on the plausible, but not formalized assumption that they provide outlier-resistance regardless of the outliers being scattered symmetrically around the center or not.

\item
An absent edge in a Gaussian graphical model, i.~e., a zero partial correlation, has the interpretation of conditional independence, a notion we have repeatedly used in Section \ref{sec:mpa}. Under ellipticity, an absent edge has the slightly weaker interpretation of conditional uncorrelatedness. This is occasionally mentioned as a limitation of elliptical graphical models. However, this limitation is largely void. 
The conclusions of any statistical analysis are always a combination of the information contained in the data and the modeling assumptions. 
When performing a purely linear analysis based on partial correlations, concluding conditional independences due the normality assumption certainly falls into the latter category.

\end{enumerate}

\section*{Technical Appendix}

The normalizing constant in the elliptical $t_{\nu}$-density is 
\[
  c_{\nu,p} =  \frac{\Gamma\{(\nu + p)/2\}}{ (\nu \pi)^{p/2} \Gamma(\nu/2)},
\]
where $\Gamma$ is the gamma function, see e.~g. \citet[Chapter 13]{Bilodeau1999}.

The Kronecker product $A \otimes B$ of two matrices $A,B \in \R^{p \times p}$ is defined as the $p^2 \times p^2$ matrix with entry $a_{i,j} b_{k,l}$ at position $((i-1)p + k, (j-1)p + l)$. For a matrix $A = (\ba_1, \ldots, \ba_p) \in \R^{p \times p}$, the notation $\vv(A)$ means the $p^2$-vector obtained by stacking the columns of $A$, i.~e., $\vv(A) = (\ba_1^\top, \ldots, \ba_p^\top)^\top$. 
The matrix $\mM_p$ in (\ref{eq:Mp}) is defined as
\[	   
  \mM_p = \frac{1}{2}\left( \mI_{p^2} + \mK_p \right),
\]
where $\mI_{p^2}$ denotes the $p^2 \times p^2$ identity matrix and
\[
	  \mK_p = \sum\nolimits_{i=1}^p \sum\nolimits_{j=1}^p \be_i^{} \be_j^\top \otimes \be_j^{} \be_i^\top,
\]
where $\be_1, \ldots, \be_p$ denote the Euclidean basis vectors in $\R^p$. 
The matrix $\mK_p$, commonly referred to as the \emph{commutation matrix}, is orthogonal and corresponds to the transpose operator  $\mK_p \vv (A) = \vv (A^\top)$. The idempotent matrix $\mM_p$ is called the \emph{symmetrization matrix} since it maps $\vv(A)$ to $\vv(A + A^\top)/2$.

\section*{Acknowledgment}

This article appears in the festschrift for David E.~Tyler. We thank the anonymous referees for many valuable suggestions and spotting mistakes in the originally submitted manuscript.


\bibliographystyle{apalike}


\end{document}